# The Aharonov-Casher theorem and the axial anomaly in the Aharonov-Bohm potential

Alexander Moroz*

*Division de Physique Théorique,[†] Institut de Physique Nucléaire,
Université Paris-Sud, F-91 406 Orsay Cedex, France*
*and*
[‡]*School of Physics and Space Research, University of Birmingham, Edgbaston,
Birmingham B15 2TT, U. K.*

## ABSTRACT

The spectral properties of the Dirac Hamiltonian in the the Aharonov-Bohm potential are discussed. By using the Krein-Friedel formula, the density of states (DOS) for different self-adjoint extensions is calculated. As in the nonrelativistic case, whenever a bound state is present in the spectrum it is always accompanied by a (anti)resonance at the energy. The Aharonov-Casher theorem must be corrected for singular field configurations. There are no zero (threshold) modes in the Aharonov-Bohm potential. For our choice of the 2d Dirac Hamiltonian, the phase-shift flip is shown to occur at only positive energies. This flip gives rise to a surplus of the DOS at the lower threshold coming entirely from the continuous part of the spectrum. The results are applied to several physical quantities: the total energy, induced fermion-number, and the axial anomaly.

---

*e-mail address : moroz@fzu.cz, am@th.ph.bham.ac.uk
[†]Unité de Recherche des Universités Paris XI et Paris VI associée au CNRS
[‡]Present address

**1. *Introduction*.-** In this paper the change $\triangle\rho_\alpha(E)$ over all space of the density of states (DOS) induced by the Aharonov-Bohm (AB) potential $\mathbf{A}(r)$ [1],

$$A_r = 0, \qquad A_\varphi = \frac{\Phi}{2\pi r} = \frac{\alpha}{2\pi r}\Phi_0, \tag{1}$$

in the radial gauge, is calculated for the two-dimensional (2d) Dirac Hamiltonian. The AB potential can be uderstood in a more general sense here. Usually $\Phi = \alpha\Phi_0$ is the total flux through the flux tube and $\Phi_0$ is the flux quantum, $\Phi_0 = hc/|e|$. However, the same potential (1) is induced by a cosmic string provided the identification $\alpha = e/Q_{Higgs}$ is made, where $e$ and $Q_{Higgs}$ are respectively the charge of a test particle and the charge of the Higgs particle [2, 3]. In what follows, $n$ and $\eta$ will be respectively the integer and the fractional part of $\alpha = n + \eta$.

If one neglects anomalous magnetic moment of the electron, the Dirac Hamiltonian in the AB potential (and in any magnetic field) considered as an operator in $L^2$ is a formal operator with supersymmetry [4]. A contradiction between periodicity of the spectrum under the substitution $\alpha \to \alpha \pm 1$ (as long as $\alpha$ does not change its sign) and the Aharonov-Casher (AC) theorem [5] is resolved. One finds that there are no threshold (zero) modes in the AB potential for any $\alpha$ (see Fig. 1). This can be confirmed independently by the calculation of the Witten index [6]. Similarly as in the nonrelativistic case [7], our calculation of the contribution of continuous spectrum to the integrated DOS (IDOS) will be based on the analysis of phase shifts and use of the Krein-Friedel formula [8, 9] regularized by the $\zeta$-function [7]. First, we shall consider a generic self-adjoint extension, when the supersymmetry of the Dirac Hamiltonian is lost and a bound state with energy $E_b$ occurs in the spectrum in the channel $l = -n - 1$. We shall calculate the change of the IDOS and show that the bound state is always accompanied by a (anti)resonance [10]. In the supersymmetric case, after taking the limits $E_b \to \pm m$ in the channel $l = -n - 1$, the bound state disappear and instead of the resonance the phase-shift flip (cf. [11]) occurs.

The DOS provides an important link between different physical quantities. The knowledge of $\triangle\rho_\alpha(E)$ determines the spectral asymmetry $\sigma_\alpha(\mathcal{E})$,

$$\sigma_\alpha(\mathcal{E}) = \triangle\rho_\alpha(\mathcal{E}) - \triangle\rho_\alpha(-\mathcal{E}), \tag{2}$$

where $\mathcal{E} = |E| > 0$. In contrast to three dimensions, the spectrum of the 2d Dirac Hamiltonian is in general *asymmetric* and, if supersymmetry holds, the spectral asymmetry is concentrated at the thresholds. The spectral asymmetry measures how the vacuum is distorted under the presence of an external field, and therefore it is natural that spectral asymmetry $\sigma_\alpha(\mathcal{E})$ determines the one-loop contribution $E^1_{eff}$ to the effective energy, the induced fermion number $Q$, and, in the case of the massless Euclidean Dirac operator, the axial anomaly $\mathcal{A}$. One has

$$E^1_{eff} = \int_0^\infty \mathcal{E}\, \sigma_\alpha(\mathcal{E})\, d\mathcal{E}, \quad Q = -\frac{1}{2}\int_0^\infty \sigma_\alpha(\mathcal{E})\, d\mathcal{E}, \quad \mathcal{A} = \int_0^\infty \sigma_\alpha(\mathcal{E})\, d\mathcal{E, } \tag{3}$$

and $Q = -\mathcal{A}/2$ [12, 13]. By using the above relations one can check for the consistency of the result for either of these quantities.

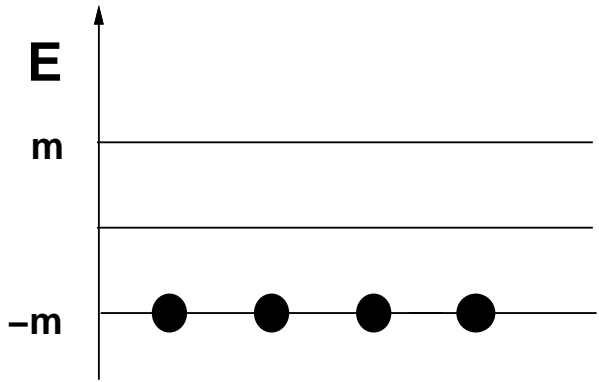

Figure 1: In the case of a 2d Dirac Hamiltonian in a generic finite-flux magnetic field the threshold states occur at one and only one threshold. There are, however, no threshold (zero) modes in the AB potential. In the latter case, the spectral asymmetry is entirely accounted for by the continuous part of the spectrum. The phase-shift flip which occurs at positive energies in the $l = -n - 1$ channel gives rise to the net surplus of $\eta$ states at the lower threshold.

2. *The Dirac Hamiltonian in the AB potential.-* The Dirac Hamiltonian in a magnetic field (in units with $\hbar = c = 1$) is

$$H(A) = \begin{pmatrix} m & D^\dagger \\ D & -m \end{pmatrix}, \quad D = \sum_{j=1}^{2}(p_1 + A_1) + i(p_2 + A_2), \tag{4}$$

where $p_j = i\partial_j$. We have taken the standard representation of $\gamma$ matrices in 2d, $\gamma^0 = \beta = \sigma_3$, $\gamma^1 = i\sigma_2$, and $\gamma^2 = -i\sigma_1$, where $\sigma_j$ are the Pauli matrices. In what follows, the charge $e$ will be assumed to be negative, $e = -|e|$.

In the AB potential, after separation of variables in polar coordinates, $H(A)$ reduces to the direct sum, $H(A) = \oplus_l h_{m,l}$, of channel operators $h_{m,l}$ in $L^2[(0,\infty), rdr]$,

$$h_{m,l} = \begin{bmatrix} m & -i\left(\partial_r + \frac{\nu+1}{r}\right) \\ -i\left(\partial_r - \frac{\nu}{r}\right) & -m \end{bmatrix}, \tag{5}$$

where $\nu = l + \alpha$ [3]. One finds immediately that $h^*_{m,l}|_{m \to -m} = -h_{m,l}$, where '$*$' means the complex conjugation. Therefore, for $E = -\mathcal{E} < -m$, the scattering states are given by

$$\Psi_{-\mathcal{E};l}(t,r,\varphi) = \Psi^*_{\mathcal{E};l}(t,r,\varphi)|_{m \to -m}. \tag{6}$$

**2.1.** *Point spectrum and the Aharonov-Casher theorem.*- Formally, the set (5) of radial Hamiltonians is invariant under the substitution $\alpha \to \alpha \pm 1$. As long as $\alpha$ does not change its sign, the spectrum of $H(A)$ must be periodic in $\alpha$ (see below). This is, however, seemingly in contradiction with the AC theorem [5] which states that the point spectrum is a function of $\alpha$. In a general finite-flux magnetic field $B(\mathbf{r})$,

$$\int_\Omega B(\mathbf{r}) \, d^2\mathbf{r} = \Phi = \text{const}, \tag{7}$$

the AC theorem tells us that $H(A)$ has exactly either $\alpha - 1 = n - 1$ or $n$ threshold states at one and only one of the thresholds $E = \pm m$ (in the present case at $E = -m$) depending whether the flux $\alpha$ is an integer or not. In the case of the 2d massless Euclidean Dirac operator, the threshold states are actually zero modes.

To resolve this contradiction, one must check for validity of the AC theorem in the case of a singular flux tube with zero radius. In the latter case, one must check for square integrability not only at infinity but at the position of singularities of the field, too. It is here where the theorems fail. If one takes proof of the AC theorem as presented, for example in Ref. [4], p. 198, and substitutes for $B(\mathbf{r})$,

$$B(\mathbf{r}) = 2\pi\alpha\,\delta^{(2)}(\mathbf{r}) = \frac{\alpha}{r}\delta(r), \tag{8}$$

one finds immediately that threshold (zero) modes are

$$e^{-\phi(r)},\ e^{-\phi(r)}(x_1 - ix_2),\ \ldots,\ e^{-\phi(r)}(x_1 - ix_2)^{n-1}, \tag{9}$$

where

$$\phi(\mathbf{r}) = \frac{e}{2\pi}\int_{R^2} \ln|\mathbf{r} - \mathbf{r}'|\,B(\mathbf{r}')\,d^2\mathbf{r}' = \alpha \ln|\mathbf{r}|. \tag{10}$$

Threshold (zero) modes are obviously singular at the origin and are not elements of $L^2[(0,\infty), rdr]$. One can show this directly by using $h_{m,l}$ as well. For a threshold state at $E = m$ to exist, the upper component $\chi_1$ of the Dirac spinor has to obey

$$\left(\partial_r - \frac{\nu}{r}\right)\chi_1(r) = 0. \tag{11}$$

At the threshold $E = -m$ one then obtains

$$\left(\partial_r + \frac{\nu+1}{r}\right)\chi_2(r) = 0. \tag{12}$$

These two equations can be easily integrated. The solutions are

$$\chi_1(r) = r^\nu \quad \text{and} \quad \chi_2(r) = r^{-(1+\nu)}, \tag{13}$$

and they are obviously not in $L^2[(0,\infty), rdr]$ for any $l$. If they are square integrable at the infinity they are not so at the origin and *vice versa*.

**2.2.** *Continuous spectrum.-* In order to use the Krein-Friedel formula [8] for the continuous part of the spectrum one must determine phase shifts. Scattering states $\Psi_{\mathcal{E},l} = \chi(r)e^{il\varphi}e^{-i\mathcal{E}t/\hbar}$ for $E > m$ are given in terms of Bessel functions with

$$\chi(r) = \frac{1}{N}\begin{pmatrix} \sqrt{\mathcal{E}+m}\,(\varepsilon_l)^l J_{\varepsilon_l\nu}(kr) \\ i\sqrt{\mathcal{E}-m}\,(\varepsilon_l)^{l+1} J_{\varepsilon_l(\nu+1)}(kr)e^{i\varphi} \end{pmatrix}, \tag{14}$$

$N$ being a normalization factor, and $\varepsilon_l = \pm 1$ [3]. The sign of $\varepsilon_l$, except for the channel $l = -n - 1$, is fixed by the square integrability condition at the origin [3]. The two-component solutions of the massive Dirac equation have only *one degree of freedom* that is reflected in the equality of up and down phase shifts [14]. Except for the channel $l = -n-1$, phase shifts are invariant under the change of the sign of the energy, $E \to -E$, and are given by

$$\delta_l^u = \delta_l^d = \begin{cases} -\pi\alpha, & l > -n-1, \\ \pi\alpha, & l < -n-1. \end{cases} \tag{15}$$

In the channel $l = -n - 1$ *two different solutions* are possible,

$$\chi^-(r) = \frac{1}{N}\begin{pmatrix} \sqrt{\mathcal{E}+m}\,J_{\eta-1}(kr) \\ i\sqrt{\mathcal{E}-m}\,J_\eta(kr)e^{i\varphi} \end{pmatrix}, \tag{16}$$

and

$$\chi^+(r) = \frac{1}{N}\begin{pmatrix} \sqrt{\mathcal{E}+m}\,J_{1-\eta}(kr) \\ -i\sqrt{\mathcal{E}-m}\,J_{-\eta}(kr)e^{i\varphi} \end{pmatrix}, \tag{17}$$

with *opposite* phase shifts,

$$\delta^- = -\frac{1}{2}\pi\eta, \qquad\qquad \delta^+ = \frac{1}{2}\pi\eta. \qquad (18)$$

Note that state $\chi^+$ ($\chi^-$) is *regular* (singular) in the limit $\eta \to 0_+$ and *singular* (regular) in the limit $\eta \to 1_-$. Provided that one takes solution $\chi^+$ for $l = -n - 1$, then phase shifts for all $l$ are given by $\delta_l = \frac{1}{2}\pi(|l| - |l + \alpha|)$. This ambiguity in the channel $l = -n - 1$ is the consequence of the fact that the AB potential is in the so-called *limit circle case* at the origin (see Ref. [15], p.152) where the boundary conditions have to be specified. A given combination of $\chi^+$ and $\chi^-$ then characterizes a particular self-adjoint extension of $H(A)$ [3]. Provided that $\alpha = -n - \eta < 0$, the critical channel is $l = n$. Therefore, despite the fact that the set (5) of radial Hamiltonians is invariant under the substitution $\alpha \to \alpha \pm 1$, the spectrum is not. The spectrum is periodic only as long as $\alpha$ does not change its sign.

3. *Self-adjoint extensions.*- Distinguished feature of a generic self-adjoint extension is that the point spectrum is not empty and a bound state can occur with energy inside the mass gap $(-m, m)$ [3, 16]. Both supersymmetry and the scale invariance are broken in this case. The bound state of energy $E_b$ in the channel $l = -n - 1$ is given in terms of modified Bessel functions $K_\nu(z)$ [17],

$$B_{E_b; -n-1}(t, r, \varphi) = \frac{1}{N} \begin{pmatrix} \sqrt{m + E_b}\, K_{1-\eta}(\kappa r) \\ i\sqrt{m - E_b}\, K_\eta(\kappa r) e^{i\varphi} \end{pmatrix} e^{-i(n+1)\varphi} e^{-iE_b t/\hbar}, \qquad (19)$$

where $\kappa = \sqrt{m^2 - E_b^2}$. In the presence of a bound state of energy $E_b$, scattering states with energy $E > m$ take a general form

$$\Psi_{\mathcal{E}; -n-1}(t, r, \varphi) = \left[\chi^+(r)\sin\mu + \chi^-(r)\cos\mu\right] e^{-i(n+1)\varphi} e^{-i\mathcal{E}t/\hbar}, \qquad (20)$$

where $\chi^+(r)$ and $\chi^-(r)$ are given respectively by (16) and (17). In contrast to [3], we shall not use the factor $(-1)^{n+1}$ for $\chi^-(r)$. Such as in the non-relativistic case [7], in the presence of a bound state of energy $E_b$ the parameter $\mu$ of the self-adjoint extension is a function of energy $E$,

$$\tan\mu = \frac{m - E_b}{(m^2 - E_b^2)^\eta} \frac{(\mathcal{E}^2 - m^2)^\eta}{\mathcal{E} - m}. \qquad (21)$$

The formula is only valid for $E > m$. Scattering states at energies $E < -m$ are obtained directly from positive energy states by using relation (6). If necessary, $\mu_\pm$ will be written to distinguish between positive and negative energies. A close inspection of (21) reveals that under $m \to -m$ both the sign and the value of $\tan \mu_+$ change,

$$\tan \mu_+ \to \tan \mu_- = -\frac{m + E_b}{(m^2 - E_b^2)^\eta} \frac{(\mathcal{E}^2 - m^2)^\eta}{\mathcal{E} + m}. \tag{22}$$

The change amounts to the following two changes in (21), $E_b \to -E_b$, and $\mathcal{E} \to -\mathcal{E}$. The corresponding phase shift is given by

$$\delta_{-n-1}(E) = \frac{1}{2}\pi\alpha + \triangle_{-n-1}(E), \tag{23}$$

where

$$\triangle_{-n-1} = -\arctan \frac{\sin(\eta\pi)}{\cos(\eta\pi) - \tan \mu}. \tag{24}$$

The analysis presented here again shows consistency of our result about the absence of threshold (zero) modes in the AB potential. As we have seen, the $l = -n - 1$ channel is the only channel where the sign of factor $\varepsilon_l$ in (14) is not fixed by the square integrability condition at the origin and where, in principle, a bound state can appear. On the other hand, the AC theorem allows for zero modes only in the channels $l = 0, -1, \ldots, -n + 1$ [cf. (9)], none of which is the critical channel $l = -n - 1$.

**4.** *The density of states and the Krein-Friedel formula*.- Let us denote by $N_V(E)$ and $N_0(E)$ respectively the IDOS induced by the presence of a scatterer $V$ and the free IDOS. The contribution of scattering states to the change $\triangle N_V(E)$ of the IDOS induced by the presence of a scatterer with a finite range of interaction is known to be determined by the Krein-Friedel formula [8] directly as the sum over all phase shifts,

$$\triangle N_V(E) = N_V(E) - N_0(E) = \frac{1}{\pi} \sum_l \delta_l(E) = (2\pi i)^{-1} \ln \det S, \tag{25}$$

S being the total on-shell S-matrix. In a recent papers [7] we have shown that the Krein-Friedel formula can have a wider range of applicability. In particular, we have established that the Krein-Friedel formula can also be used for the long-ranged AB potential, provided that the sum over phase shifts is regularized by the $\zeta$-function. The fact that phase shifts can be rather easily calculated without any care of the proper normalization of wave functions greatly facilitates the calculation. Moreover, by means of the Krein-Friedel formula it is rather easy to calculate the change of the IDOS for all possible self-adjoint extensions of (4). In the AB potential, the IDOS $N_\alpha(E)$ is as usually defined by $N_\alpha(E) =$ $\int_{-\infty}^{E} \rho_\alpha(E')\, dE'$, where $\rho_\alpha(E) = -(1/\pi) \operatorname{Im} \operatorname{Tr} G_\alpha(\mathbf{x}, \mathbf{x}, E + i\epsilon)$, and $G_\alpha(\mathbf{x}, \mathbf{x}, E + i\epsilon)$ is the resolvent of $H(A)$ in the AB potential with the flux corresponding to $\alpha$. By using the $\zeta$-function regularization [7] one finds in each energy continuum that

$$
\begin{aligned}
\ln \det S &= \sum_{l=-\infty}^{\infty} 2i\delta_l = 2i\triangle_{-n-1} + i\pi \sum_{l=-\infty}^{\infty} (|l| - |l + \alpha|) \\
&= 2i\triangle_{-n-1} + i\pi \left[ 2\sum_{l=1}^{\infty} l^{-s} - \sum_{l=0}^{\infty}(l+\eta)^{-s} - \sum_{l=1}^{\infty}(l-\eta)^{-s} \right]\bigg|_{s=-1} \\
&= 2i\triangle_{-n-1} + i\pi \left[ 2\zeta_R(s) - \zeta_H(s, \eta) - \zeta_H(s, 1 - \eta) \right]\big|_{s=-1} \\
&= 2i\triangle_{-n-1} - i\pi \eta (1 - \eta),
\end{aligned} \quad (26)
$$

where $\zeta_R$ and $\zeta_H$ are the Riemann and the Hurwitz $\zeta$-functions. Therefore, according to the Krein-Friedel formula (25),

$$
\triangle N_\alpha(E) = -\frac{1}{2}\eta(1-\eta) - \frac{1}{\pi} \arctan \frac{\sin(\eta\pi)}{\cos(\eta\pi) - \tan\mu}. \tag{27}
$$

One can repeat the calculation for $\alpha \leq 0$ to verify that

$$
\triangle N_{-|\alpha|}(-E) = \triangle N_{|\alpha|}(E). \tag{28}
$$

**4.1.** *The resonance.*- $\tan\mu_+$ is positive for $E > m$ [see (21)] and hence there is a typical *(anti)resonance* in the relativistic case for $0 < \eta < 1/2$ at the energy $\mathcal{E}_{res}$ which satisfies

$$
\frac{\mathcal{E}_{res} - m}{(\mathcal{E}_{res}^2 - m^2)^\eta} = \frac{1}{\cos\eta\pi} \frac{m - E_b}{(m^2 - E_b^2)^\eta}. \tag{29}
$$

The phase shift $\delta_{-n-1}(E)$ (23) changes by $-\pi$ in the direction of increasing energy and the integrated DOS (27) has a sharp *decrease* by one. For $1/2 < \eta < 1$ the cosine in (27) changes its sign and the antiresonance disappears at positive energies [18]. Instead, a resonance occurs at the negative energy continuum at the energy $E_{res} = -\mathcal{E}_{res}$ which is determined by the equation

$$
\frac{\mathcal{E}_{res} + m}{(\mathcal{E}_{res}^2 - m^2)^\eta} = \frac{1}{|\cos\eta\pi|} \frac{m + E_b}{(m^2 - E_b^2)^\eta}. \tag{30}
$$

$\eta = 1/2$ is a special point since the (anti)resonance is at infinity. Obviously, the shape of the (anti)resonance is not of the Breit-Wigner type.

**4.2.** *Supersymmetric case and the phase-shift flip.*- One knows that in the supersymmetric case there is neither bound state nor the (anti)resonance. Therefore, to get rid of the bound state we shall consider limits $E_b \to \pm m$. If $E_b$ changes, so do $E_{res}$ and $\tan \mu_\pm$. Limiting values of $E_{res}$ and $\tan \mu_\pm$ are presented in Tab. I.

TABLE I. Limiting values of $E_{res}$ and $\tan \mu_\pm$ when $E_b \to \pm m$.

|  | $0 < \eta < 1/2$ | $1/2 < \eta < 1$ | $0 < \eta < 1/2$ | $1/2 < \eta < 1$ |
|---|---|---|---|---|
| $E_b \uparrow m$ | $E_{res} \downarrow m$ | $E_{res} \uparrow -m$ | $\tan \mu_+ \to 0$ | $\tan \mu_- \to -\infty$ |
| $E_b \downarrow -m$ | $E_{res} \uparrow \infty$ | $E_{res} \downarrow -\infty$ | $\tan \mu_+ \to \infty$ | $\tan \mu_- \to 0$ |

Formally, both cases $E_b \uparrow m$ and $E_b \downarrow -m$ seem to be plausible. In each case, scattering states in one of the energy continuums are described by state $\chi^+$ and in the other continuum by state $\chi^-$. As a result, one has the *phase-shift flip* in one energy continuum,

$$\delta^+ \to \delta^- = -\frac{1}{2}\pi\eta,$$

and the phase shift $\delta^+$ in the second energy continuum. In order that our results be compatible with previous calculations of the spectral aymmetry and the axial anomaly [13], the limit $E_b \uparrow m$ was chosen. In this case, the phase-shift flip occurs at positive energy continuum while at the negative energy continuum the phase shift is $\delta^+ = \pi\eta/2$. Therefore, the contribution of scattering states with energy $E \geq m$ to the change of the DOS is

$$\triangle\rho_\alpha(E) = \rho_\alpha(E) - \rho_0(E) = -\frac{1}{2}\eta(3-\eta)\,\delta(E-m). \tag{31}$$

while the contribution of scattering states with $E \leq -m$ to the change of the DOS is

$$\triangle\rho_\alpha(E) = -\frac{1}{2}\eta(1-\eta)\,\delta(E+m). \tag{32}$$

As expected, the change of the DOS is concentrated at the thresholds where it is proportional to delta functions.

5. *Spectral asymmetry, effective energy, induced fermion number, and the axial anomaly.-* In the absence of the bound state, when supersymmetry holds, the spectral asymmetry is entirely determined by the continuous part of the spectrum. One finds from (31) and (32) that the contribution is actually determined by the phase-shift flip,

$$\sigma_\alpha(\mathcal{E}) = -\eta\,\delta(\mathcal{E} - m). \tag{33}$$

Correspondingly, according to (3),

$$E^1_{eff} = -m\eta, \qquad Q = \frac{1}{2}\eta, \qquad \mathcal{A} = -\eta. \tag{34}$$

The phase-shift flip which occurs at positive energies in the $l = -n - 1$ channel gives rise to the net surplus of $\eta$ states at the lower threshold with respect to the upper energy threshold (see Fig. 1). When $\alpha$ reaches 1, then, in the limit $\alpha \to 1_-$, there is seemingly a surplus of one state at the lower threshold with respect to the upper threshold. However, for $\alpha = 1$ all quantities of interest, including basic set (5) of radial Hamiltonians and wave functions (14) are identical to those in the case for $\alpha = 0$. By using the fact that phase shifts are only defined modulo $\pi$, one finds that phase shifts depend periodically (although discontinuously) on $\alpha$ [cf. (15)]. Therefore, we argue that spectral asymmetry $\sigma_\alpha$, and consequently $E^1_{eff}$, $Q$, and $\mathcal{A}$ change *discontinuously* to their original zero values for $\alpha = 1$. This can already be seen from our results (31-32) for the change of the DOS. Actually, we have already encountered a discontinuous behaviour in the discussion of the phase-shift flip. Despite that $\chi^-$ is singular in the limit $\eta \to 0$, it is this state which occurs in the spectrum for $E \geq m$, and not state $\chi^+$.

Another justification for our conclusion for the discontinuous behaviour of the spectral asymmetry comes from the nonrelativistic case. In the latter case one knows that there exists one physical quantity closely associated with the DOS, and hence with the spectral asymmetry, which changes discontinuously with the flux $\alpha$ [19]. This experimentally measurable quantity [20] is the so-called *persistent current* induced by the AB potential [21].

In the case for a generic self-adjoint extension when the bound state (19) is present, the supersymmetry is lost, because spectral asymmetry $\sigma_\alpha$ is no longer concentrated at the mass thresholds, and respective restrictions of $D$ on $(\text{Ker}\, D)^\perp$ and $D^\dagger$ on $(\text{Ker}\, D^\dagger)^\perp$ are no longer unitary equivalent. By using results (27) for the IDOS one finds

$$\sigma_\alpha(\mathcal{E}) = \text{sign}(E_b)\,\delta(\mathcal{E} - |E_b|) + \frac{1}{\pi}\frac{d}{d\mathcal{E}}\left\{\left[\triangle^+_{-n-1}(\mathcal{E}) - \triangle^-_{-n-1}(\mathcal{E})\right]\Theta(\mathcal{E} - m)\right\}, \tag{35}$$

where $\Theta$ is the Heaviside step function, and $\triangle^{\pm}_{-n-1}(\mathcal{E})$ is the change of the conventional phase shift (24) respectively for positive and negative energy states in the $l = -n - 1$ channel. The superscripts '$\pm$' means that the value of $\tan \mu_{\pm}$ is taken in (24). The values of $E^1_{eff}$, $Q$, and $\mathcal{A}$ are then obtained by substituting (35) in (3). It should be stressed out that $\sigma_\alpha(\mathcal{E})$ can have both signs depending on the sign of the mass in a 2d Dirac Hamiltonian.

**6.** *Discussion of the results.-* By extending the Krein-Friedel formula to the case of a singular Aharonov-Bohm (AB) potential we were able to express all physical quantities discussed here in terms of scattering phase shifts. We have

- obtained closed analytical form for the density of states $\rho_\alpha(E)$ of the 2d Dirac Hamiltonian induced by the AB potential and predicted a resonance to occur whenever a bound is in the spectrum;

- calculated spectral asymmetry $\sigma_\alpha(\mathcal{E})$, the one-loop contribution $E^1_{eff}$ to the effective energy, the induced charge $Q$, and the axial anomaly $\mathcal{A}$ (cf. [22]) of the Dirac operator in the AB potential for a generic self-adjoint extension;

- pointed out that the Aharonov-Casher theorem has to be corrected for singular field configurations;

- determined that the phase-shift flip only occurs for positive energies.

In a particular case, when the bound state is absent and supersymmetry holds, our result for $\triangle \rho_\alpha(E)$ are consistent with an earlier calculations of the spectral asymmetry $\sigma_\alpha(\mathcal{E})$, $Q$, and $\mathcal{A}$ [13]. Relation between the axial anomaly $\mathcal{A}$ and the phase-shift flip has been discussed in [23]. Their treatment, however, differs from ours since they considered a complementary situation of a regular finite-flux configuration and the limit $k \to 0$. The phase-shift flip cannot take place for both positive and negative energy scattering states (cf. Ref. [11]). One of the arguments is that in the latter case the spectral asymmetry, and correspondingly induced charge and the axial anomaly all would be zero. Physically, a bound state can occur in the spectrum when an attractive $\delta$-function potential is put on top of the AB potential [16, 24]. It this case both supersymmetry and scale invariance are broken, the differential scattering cross section is asymmetric, and the Hall effect occur [24].

Note that one has the unitary equivalence between a spin 1/2 charged particle in a 2d magnetic field and a spin 1/2 neutral particle with an anomalous magnetic moment in a 2d electric field [25] and our results apply to the this case as well. An application of our results to persistent currents in the field of a cosmic string [2, 3] and in the field of a gravitational vortex, and the discussion of the total energy in the AB potential are given elsewhere [24].

I should like to thank A. Comtet, Y. Georgelin, M. Knecht, S. Ouvry, and J. Stern for many useful and stimulating discussions, and R. C. Jones for careful reading of the manuscript.